\documentclass[fleqn,usenatbib]{mnras}

\usepackage{booktabs,caption}
\usepackage{newtxtext,newtxmath}
\usepackage[flushleft]{threeparttable}

\usepackage[T1]{fontenc}

\DeclareRobustCommand{\VAN}[3]{#2}
\let\VANthebibliography\thebibliography
\def\thebibliography{\DeclareRobustCommand{\VAN}[3]{##3}\VANthebibliography}

\usepackage{graphicx}	
\usepackage[dvipsnames]{xcolor}

\newcommand{\kms} {\ifmmode{\rm \,km\,s^{-1}}\else\,km\,s$^{-1}$\xspace\fi}

\newcommand{\ha}{\hbox{H$\alpha$}}
\newcommand{\hb}{\hbox{H$\beta$}}

\newcommand{\nii}{\hbox{[N\,{\sc ii}]}}
\newcommand{\oiii}{\hbox{[O\,{\sc iii}]}}

\title[One-sided \ha\ Excess in Galaxy Pairs]{One-sided \ha\ Excess before the First Pericentre Passage in Galaxy Pairs}
\author[Jiwon Chung et al.]{
Jiwon Chung,$^{1}$\thanks{E-mail: jiwon@kasi.re.kr}
Joon Hyeop Lee,$^{1}$
Hyunjin Jeong $^{1}$
\\
$^{1}$Korea Astronomy and Space Science Institute, Republic of Korea\\
}

\date{Accepted June 4. Received 2024 June 4; in original form 2024 March 4}

\pubyear{2015}

\begin{document}
\label{firstpage}
\pagerange{\pageref{firstpage}--\pageref{lastpage}}
\maketitle

\begin{abstract}
We present novel insights into the interplay between tidal forces and star formation in interacting galaxies before their first pericentre passage. We investigate seven close pair galaxies devoid of visible tidal disturbances, such as tails, bridges, and shells. Using integral field spectroscopy (IFS) data of extended Calar Alto Legacy Integral Field Area (eCALIFA), we unveil a previously unreported phenomenon: \ha\ emission, a proxy for recent star formation, exhibits a significant enhancement in regions facing the companion galaxy, reaching up to 1.9 times higher flux compared to opposite directions. Notably, fainter companions within pairs display a more pronounced one-sided \ha\ excess, exceeding the typical range observed in isolated galaxies with 2$\sigma$ confidence level. Furthermore, the observed \ha\ excess in fainter companion galaxies exhibits a heightened prominence at the outer galactic regions. These findings suggest that tidal forces generated before the first pericentre passage exert a stronger influence on fainter galaxies due to their shallower potential wells by their brighter companions. This unveils a more intricate interplay between gravitational interactions and star formation history within interacting galaxies than previously understood, highlighting the need further to explore the early stages of interaction in galaxy evolution.

\end{abstract}


\section{Introduction}

Numerous studies have concluded that gravitational interactions between galaxies, particularly mergers, lead to temporarily increased star formation rates (\citealt{Larson78, Lonsdale84,Keel85,Barton00,Woods06,Lin07,Smith07,Ellison08}, among many others). It is well established that star formation predominantly occurs in the central regions of galaxies, irrespective of the presence or absence of tidal tails or bridge, as observed in studies of galaxy pairs \citep{Ellison10,Chown19,Pan19}. This is consistent with the predictions of simulations, which suggest that tidal forces can trigger the inflow of gas into the central regions of galaxies, leading to enhanced star formation \citep{Mihos94,Cox06,dimatteo07,Moreno15}. Low-resolution merger simulations employing the Kennicutt-Schmidt law, which parameterizes star formation rates solely as a function of local gas density and neglecting gas fragmentation, predict a highly centralized distribution of star formation within the merger remnant. These simulations suggest a suppression of star formation activity in the outskirts compared to the central region \citep{Mihos94,Mihos96,Moreno15}.

While interacting galaxies often experience nuclear star formation, they also exhibit localized episodes of star formation in their outer disks and the tidal features created by the interaction, such as warped disks and extended tails \citep{Mirabel91,Mirabel92,Hibbard96,Smith10}. \citet{Smith10} employed a comparative approach to analyze star formation rates in 46 nearby interacting galaxy pairs characterized by prominent tidal tails and bridges. This analysis contrasted these systems with 39 normal spiral galaxies. Their findings revealed that within the interacting systems, regions exhibiting the highest star formation rates occur at the points of intersection between spiral arms or tidal features. The high-resolution numerical simulations also conducted by some studies \citep{Teyssier10,Hopkins13,Renaud15,Renaud16}, capable of resolving physical processes at the parsec scale, have provided valuable insights into the star formation efficiency of galaxy mergers. These studies concluded that extended starbursts spontaneously arise in such galactic interactions due to the fragmentation of gas clouds. This fragmentation is attributed to the amplification of supersonic turbulence within the interstellar medium triggered by the tidal forces associated with the merger event.

\begin{table*}
  \begin{threeparttable}
    \caption{Basic Characteristics of the Sample Galaxies} \label{tab:table1}
     \begin{tabular}{lcccclccc}
        \toprule
	Target ID & R.A. & Decl. & $z$ & $r$ (mag.)  & Companion ID & $r$ (mag.) & Projected Separation (kpc)  & $\Delta$$v$ ($\kms$) \\
        \midrule
        
        \multicolumn{9}{c}{\centering Main Pair Sample} \\
        \hline
        UGC00312 & 7.850 & 8.467 & 0.01457 & 13.48  & UGC00312NED01 & 13.90 &24.61 & 17  \\
        UGC00312NED01 & 7.829 & 8.475 & 0.01451 & 13.90  & UGC00312 & 13.48 & 24.61 & 17  \\
		NGC0477 & 20.335 & 40.488 & 0.01960 & 13.44  & CGCG536-030 & 14.55 & 58.38 & 37   \\
        CGCG536-030 & 20.289 & 40.470 & 0.01972 & 14.55  & NGC0477 & 13.44 & 58.38 & 37   \\		
        NGC2449 & 116.835 & 26.930 & 0.01629 & 12.87  & IC0476$^*$ & 14.40 & 31.66 & 116  \\
        NGC5480 & 211.590 & 50.725 & 0.00637 & 12.25  & NGC5481$^*$ & 11.92 & 24.85 & 78  \\
        NGC3896 & 177.234 & 48.675 & 0.00307 & 13.64  & NGC3893$^*$ & 10.77 & 14.08 & 45  \\
        \hline
        
        \multicolumn{9}{c}{\centering Isolated Control Sample} \\
        \hline
        IC1256 & 260.947 & 26.487 & 0.01579 & 13.60 & - & - & - & -\\
		NGC0237  &10.866 & -0.125  & 0.01390 & 13.01  & - & - & - & -\\
		NGC5622 & 216.551 & 48.564   & 0.01289  & 13.24  & - & - & - & -\\
        NGC6063 & 241.804 & 7.979   & 0.00950 & 12.98  & - & - & - & -\\
        NGC2540	& 123.194	& 26.362   &  0.02101 & 13.06 & - & - & - & -\\
        \bottomrule
     \end{tabular}
    \begin{tablenotes}
      \small
      \item Notes: The IC 0476 is absent from the eCALIFA survey. The NGC 5481 and NGC 3893 were excluded from the main sample selection due to their non-star-forming nature or size, respectively.

    \end{tablenotes}
  \end{threeparttable}
\end{table*}

To date, the majority of numerical simulations have predominantly concentrated on the phases between after the first pericentre passage and the final coalescence, neglecting the phase before the first pericentre passage (incoming phase) during the interaction. Building upon previous results that placed the onset of enhanced star formation in galaxy interactions after the first pericentre passage, \citet{Renaud16} conducted a comprehensive simulation study. Their findings revealed that star formation could be triggered even before the first pericentre passage. This earlier initiation of star formation highlights the significant influence of tidal forces on gas dynamics within interacting galaxies, even at the early stages of the encounter. Conversely, \citet{dimatteo08} put forth an alternative perspective, suggesting that star formation activities of interacting galaxies are unchanged before the first pericentre passage. Furthermore, observational studies \citep{Pan19,Feng20} support the notion that interacting galaxies in the phase before the first pericentre passage phase exhibit integrated star formation rates that are indistinguishable from those of isolated systems. These findings further contribute to the ongoing debate regarding the precise timing of the onset of enhanced star formation in galaxy interactions. The dearth of simulations and observational studies focused on the incoming phase before the first pericentre passage highlights a critical gap in our understanding of star formation in interacting galaxies. While insights can be gleaned from numerical simulations, they cannot replicate the full complexity of galaxy interactions.

By unravelling the spatial distribution and temporal evolution of star formation activity, integral field spectroscopy (IFS) observations can offer invaluable insights into the complex interplay between tidal forces, gas dynamics, and star formation in interacting galaxies. IFS information can reveal whether star formation is triggered uniformly across the whole bodies of the pair galaxies or concentrated in specific regions, potentially influenced by the distribution and flow of gas driven by tidal forces. Moreover, the detailed spatial distribution and information on star formation age revealed by IFS data can provide crucial evidence for or against the scenarios proposed by different simulation studies. This would ultimately lead to a more comprehensive understanding of how galaxy interactions shape the evolution of both central and peripheral regions of galaxies.

In this paper, we present the spatially resolved star formation activity in the phase before the first pericentre passages using IFS data from the extended Calar Alto Legacy Integral Field Area (eCALIFA) survey \citep{Sanchez23}. The eCALIFA survey possesses a wide field-of-view (FoV) and provides access to spatially resolved galaxy data, enabling comprehensive analysis of the properties of entire galaxies. In Section 2, we describe the data and analysis of spectra based on IFS data of the eCALIFA survey. In Section 3, We perform a comparative analysis of spatially resolved \ha\ excess, considering the relative luminosity of a galaxy, parameters quantifying the tidal force by companion, and projected separation. We also discuss the results in relation to the enhancement of \ha\ and its spatial distribution in comparison with isolated galaxies. Finally, we summarize our main results in Section 4. Throughout this paper, we adopt a cosmology of $H_0$ = 70 km s$^{-1}$ Mpc$^{-1}$,  $\Omega_{m}$ = 0.3, and $\Omega_{\Lambda}$ = 0.7.

\section{Data and Analysis}
\subsection{Data}
This study analyzes the IFS datacubes provided in the eCALIFA. The eCALIFA data is obtained with the 3.5m telescope at Calar Alto Observatory, utilizing the PPAK integral field unit \citep{Kelz06} coupled with the PMAS spectrograph \citep{Roth05}. The eCALIFA employed the same V500 grating, goniometer angle, integration time (900 seconds per pointing), and dithering scheme as those conducted by the CALIFA survey \citep{Sanchez12,Sanchez16}. This configuration adopts a low-resolution setup (R$\sim$850), capturing a wavelength range of 3745 \AA\ to 7500 \AA\ within a hexagonal FoV measuring 74 $\times$ 64 arcsec$^2$. In the eCALIFA, foreground stars within the FoV are effectively masked based on the Gaia catalogue \citep{Gaia16,Gaia21}, which is crucial for investigating spatially resolved regions.

\begin{figure*}
\includegraphics[width=17cm, height=7cm]{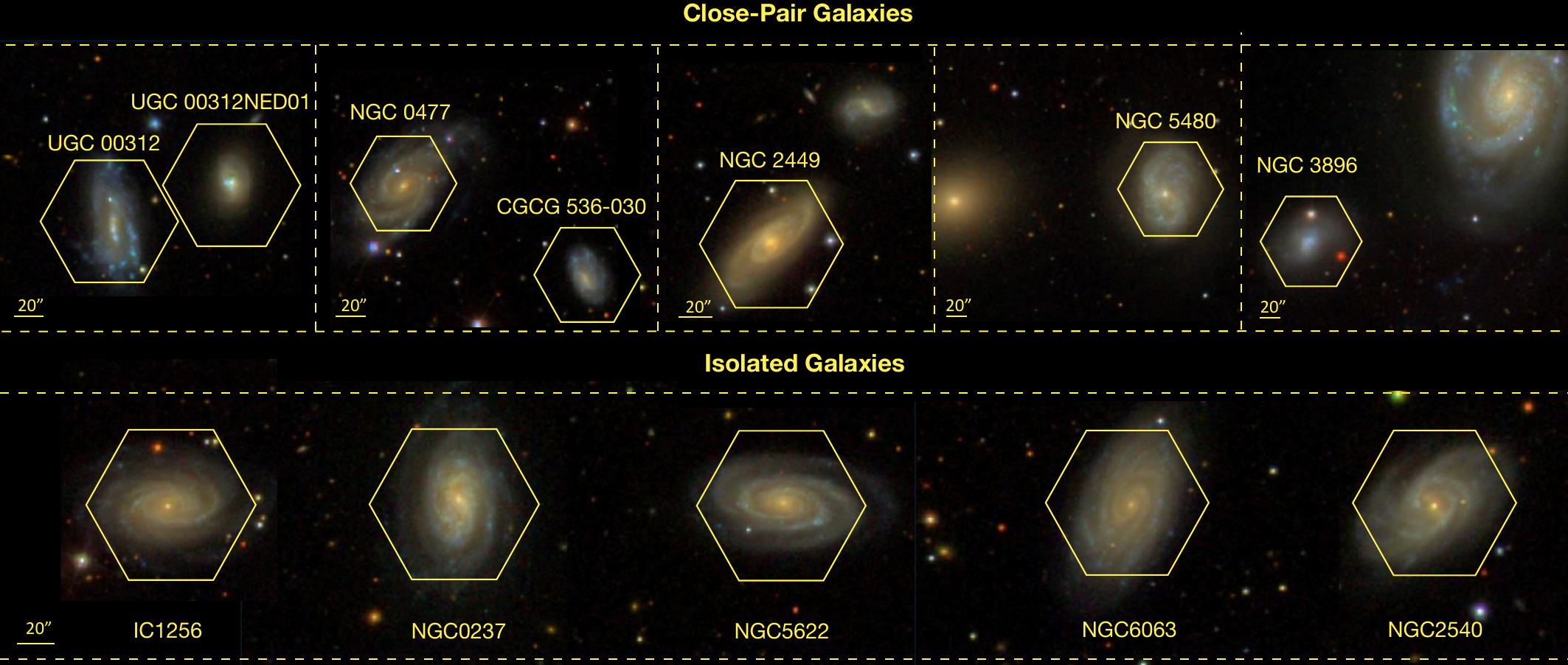}

\caption{The SDSS $g$, $r$, and $i$ combined images of main pair sample (upper) and isolated control sample (lower). The yellow hexagon indicates eCALIFA FoV. }
\label{SDSS}
\end{figure*}

\subsection{Selection of Main Close-pair Sample and Isolated Control Sample}
To investigate star formation in galaxies prior to their first pericentre passage, we employed a multi-tiered selection strategy utilizing the eCALIFA dataset. The companion galaxies utilized for the selection of the main and control samples were identified in the Sloan Digital Sky Survey (SDSS) DR12, a spectroscopic survey of galaxies with magnitude-limited down to $r$ < 17.77 \citep{Alam15}.

Firstly, we identified close pairs of galaxies based on specific criteria: a projected separation of less than 80 kpc with a relative velocity $\Delta$$v$ below 300 $\kms$, indicating likely ongoing interaction \citep{Ellison13}. To identify pair galaxies belonging to phase before the first pericentre passage, our selection criteria strictly required the absence of any discernible tidal tails or bridges \citep{Pan19,Feng20}. Secondly, we restricted our analysis to galaxies with an $r$-band Petrosian radius (R$_{90}$) smaller than 40 arcseconds in the SDSS DR12. This criterion ensures that the majority of each galaxy's structural components fall within the eCALIFA FoV, enabling comprehensive exploration of their entire properties. The galaxies for which a significant offset between the galactic centre and the eCALIFA FoV resulted in the non-observation of one side of the galactic disk were excluded from the sample. Thirdly, we refined the sample by incorporating only galaxies with an \ha\ equivalent width exceeding 5\AA\ in their central regions to investigate their star forming activity and excluded edge-on galaxies to enable robust analysis of properties across the entire galactic disk. During this process, we prioritized the elimination of early-type galaxies exhibiting emission lines potentially originating from active galactic nuclei, thereby ensuring the selection of late-type galaxies actively undergoing star formation. After careful evaluation, seven galaxies that satisfied all of the aforementioned criteria were selected. There is a possibility that the galaxy pairs selected based on projected distance may not necessarily be genuinely close pairs. However, in observational studies, there is currently no practical method to select galaxy pairs more accurately than this approach. We also note that not all companions of eCALIFA galaxies were observed in eCALIFA.

To ensure the impartiality of our results, we constructed a control sample that mirrors the selection criteria employed for the main pair sample. The key distinction between the two samples resides in the inclusion criterion for companion galaxies. The control sample strictly comprises isolated galaxies, lacking any companions within a projected separation of 500 kpc and a relative velocity of 700 $\kms$. This rigorous selection process strictly restricts the inclusion of galaxies potentially influenced by tidal interactions from nearby companions. Consequently, it facilitates a direct and unbiased comparison of the star formation properties exhibited by galaxies within interacting pairs versus those residing in isolation. Finally, five galaxies were chosen to satisfy the control sample criteria. Re-examination through visual inspection corroborated the prior finding of no galaxies within the 500 kpc radius that lack spectroscopic data from the SDSS. Figure ~\ref{SDSS} shows SDSS $g$, $r$, and $i$ composite images of the main and control samples, respectively. The basic characteristics of galaxies in main and control samples are summarized in Table ~\ref{tab:table1}.

\subsection{Measurement of Emission Line Flux and Internal Reddening Correction}

Our analysis relies on the eCALIFA data cubes employing adaptive binning, utilizing the Voronoi method \cite{Cappellari03} to achieve a continuum signal-to-noise ratio (S/N) exceeding 30 for each bin. We employed the Penalized Pixel cross-correlation Fitting (pPXF) algorithm \cite{Cappellari03} to model the stellar continuum in each bin using stellar templates from the MILES library \citep{Vazdekis10} and estimate gas components.

Measurement of emission line fluxes was conducted through a multi-step process. First, pPXF algorithm was employed to the observed spectrum, with masking applied not only to bad pixels but also to all known emission lines within a 500 $\kms$ range of each line. This step facilitated accurately determining the stellar continuum and absorption lines, which were subsequently subtracted from the observed spectrum to obtain the residual spectrum. Finally, Gaussian fitting was applied to the residual spectrum to measure the flux of each emission line.

For correction of internal extinction to the measured emission line
fluxes, a distinction between star-forming and active galactic nucleus regions is crucial. To achieve this, we employed the BPT diagram \citep{Baldwin81}, a well-established diagnostic tool relying on the emission line ratios of log($\oiii\lambda5007$/$\hb$) and log($\nii\lambda6584$/$\ha$). Then, we used the Balmer decrement of H$\alpha$/H$\beta$=2.86 and 3.1 for star-forming and AGN regions \citep{Osterbrock06}, respectively. For all bins in each galaxy, the lowest S/N of \ha\ emission line is $\sim$ 12 in all bins for the main pair galaxies used in this study.

\section{Results and Discussion}
\subsection{One-sided H$\alpha$ Flux Excess in the Pair System}

\begin{figure*}
\includegraphics[width=15cm, height=10cm]{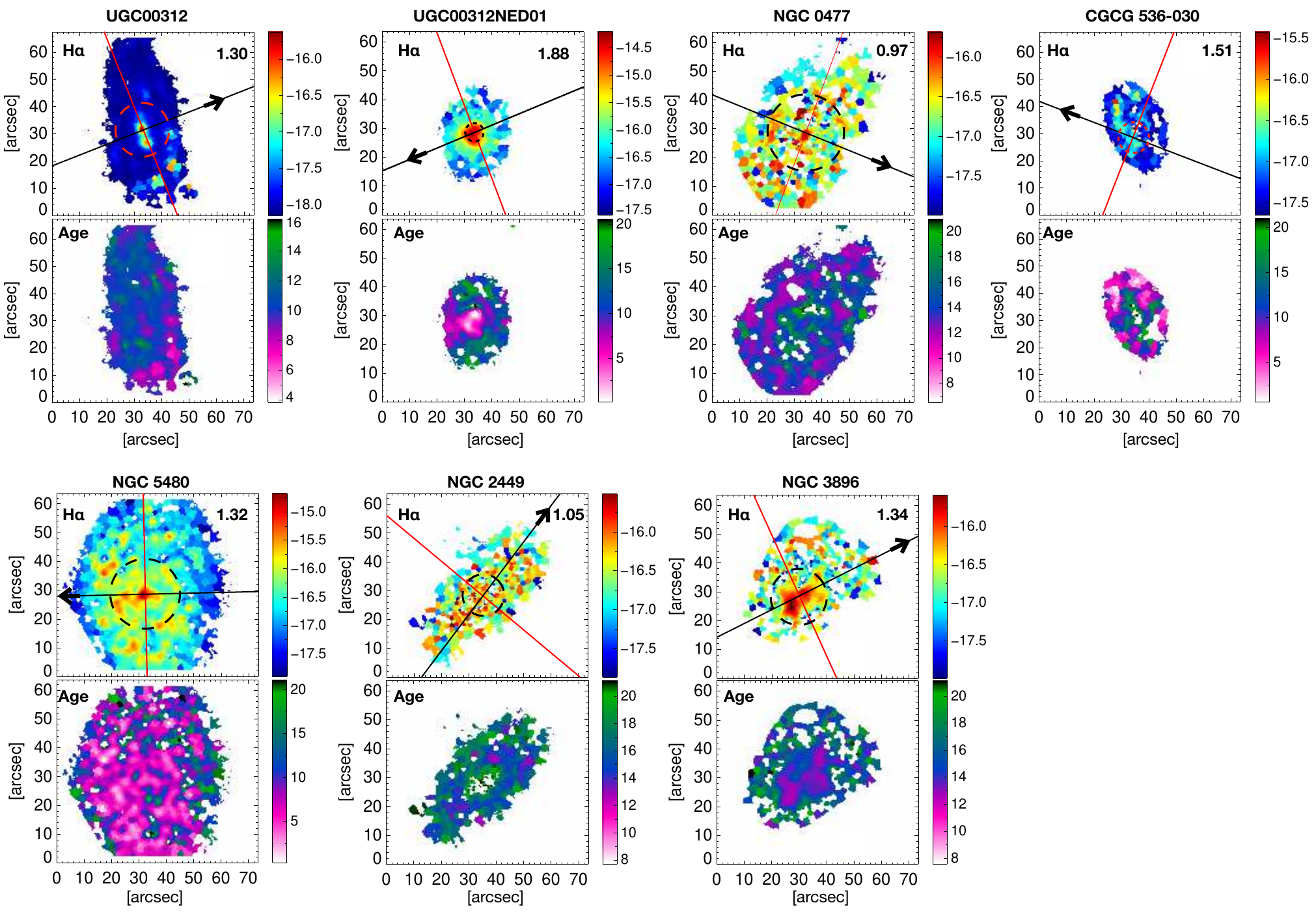}
\caption{The \ha\ flux (upper) and star formation age (bottom) maps for seven main galaxies in pairs. The units in \ha\ and age maps are (log$_{10}$\ha) erg s$^{-1}$ cm$^{-2}$ arcsec$^{-2}$ and Myr, respectively. The arrow indicates the direction of the companions. The solid black line passes through the centre of SDSS $i$-band contours of the pair galaxies. The red line is perpendicular to the black line and passes through the centre of the main galaxy. Indicated along the bottom right of each H$\alpha$ map is H$\alpha$ excess ratio. The dashed circle in the \ha\ map represents the petrosian $R$$_{50}$ radius of a galaxy.}
\label{enhance}
\end{figure*} 

\begin{figure}
\centering
\includegraphics[width=8.5cm,height=6cm]{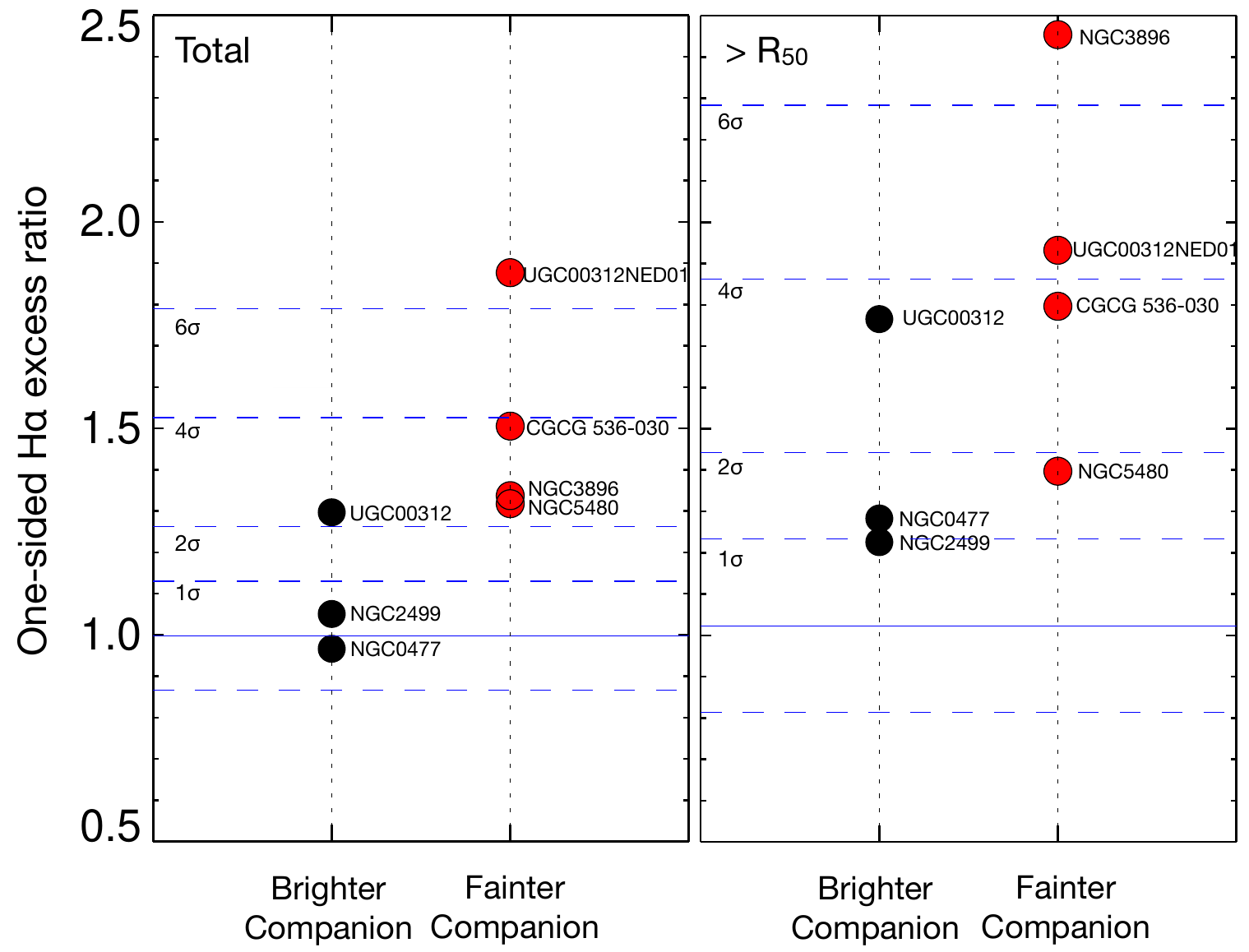}
\caption {The distributions of one-sided \ha\ excess ratios of brighter companions (black circle) and fainter companions (red circle), respectively. The left and right panels correspond to the calculations for the entire region of galaxy and the outer region of $R$$_{50}$, respectively. The solid and two blue dashed lines represent the mean, 1$\sigma$, 2$\sigma$, 4$\sigma$ and 6$\sigma$ of the \ha\ excess ratio distribution obtained by performing a random test on the isolated sample.}
\label{comp}
\end{figure}

The top of Figure ~\ref{enhance} presents the \ha\ emission line flux maps for seven main galaxies having a close companion. At first glance, it appears that the enhanced star formation regions within a galaxy are associated with the direction of the companion galaxy. We investigated whether the \ha\ flux distribution of main galaxies is correlated with the direction of the companion galaxy. We divided the main galaxies in half along the solid red line that separates the regions directly facing and opposite the companion galaxy. We then quantified the total \ha\ flux in each half and defined their ratio as the \ha\ excess ratio in the region facing the companion. This ratio serves as an indicator of the potential influence of tidal forces from the companion on one-sided star formation. The range of \ha\ excess ratio values for the seven main galaxies are about from 0.97 to 1.88. The H$\alpha$ excess ratio value is displayed in the lower right corner of each H$\alpha$ map. To estimate the age of star formation, we compared the equivalent width of the \ha\ emission line with the predictions of Starburst99 evolutionary synthesis models \citep{Leitherer99}, assuming a constant solar metallicity (Z=Z$_\odot$). We measured the average gas-phase metallicity for each main galaxy, and they fall within a range of approximately 8.4 to 9.0, while the Z$_{(O/H)}$$_\odot$=8.74 \citep{Bergemann21}. We estimated the gas-phase metallicity using the N2 ([NII]/\ha) metallicity indicator \citep{Denicolo02}. We generated models using Geneva tracks with standard mass-loss rates, assuming instantaneous star formation at 0.1 Myr timesteps with a Salpeter initial mass function and adopting the expanding atmosphere of Padrach/Hillier. Within the seven main samples, regions showing increased \ha\ excess ratio tend to be younger than other regions, with estimated star formation ages ranging from approximately 5 to 15 Myr (see bottom of Figure ~\ref{enhance}). This finding suggests that these regions have undergone recent bursts of star formation by tidal interaction.

This result is the first observational evidence that the gravitational tidal forces between close pairs might be playing a role in triggering star formation in the regions facing each other before the first pericentre passage, but the strength of this effect appears to vary from galaxy to galaxy. Numerous previous studies of galaxy interactions have primarily focused on the changes of characteristics in galaxies from after the first pericentre passage to final coalescence. This is because it was previously thought that the characteristics of galaxies before the first pericentre passage were not significantly different from those of isolated galaxies \citep{dimatteo08,Pan19,Feng20}. The physical mechanisms responsible for the one-sided \ha\ enhancement towards the companion galaxy before the first pericentre during the galaxy-galaxy interaction are still not fully understood. However, \citet{Renaud16} demonstrated through their simulation study that enhanced star formation could initiate before the first pericentre passage in all models. Notably, under the Milgromian dynamics, the lower dynamical friction facilitates a wider distribution of star formation across the galactic disk, contrasting with the Newtonian dynamics perspective, where star formation preferentially concentrates in the central regions. \

On the other hand, it is also known that star formation asymmetry can also occur in isolated galaxies by cosmological accretion of gas on galactic disks \citep{Bournaud05}. Therefore, it is necessary to examine whether the observed \ha\ excess in main galaxies can be coincidental results. Based on the five isolated galaxies, we calculated the range of \ha\ excess ratio that could be expected by chance in one direction. First, we randomly divided the isolated galaxy into two halves at 1000 different angles. We then measured the \ha\ excess ratio on one side of the galaxy in each division. This process was repeated for five isolated galaxies, resulting in measurements of \ha\ excess ratios for a total of 5000 instances. The mean and standard deviation of \ha\ excess ratio distribution of for isolated galaxies are 0.998 and 0.132, respectively, indicating a mild asymmetry of star formation.

The tidal force in galaxy interactions is known to be a significant driver of star formation during the interaction. In particular, in unequal-mass galaxy pairs, the lower-mass galaxy is more likely to experience triggered star formation than the higher-mass companion \citep{Donzelli97,Woods07,Ellison08, Li08,Alonso12} as predicted by simulations \citep{Bekki06,Cox08}. To investigate the impact of tidal forces on star formation, we divided the seven main samples into two groups based on whether each galaxy is relatively brighter or fainter than its companion galaxy in their pair system and compared the \ha\ excess for the two groups. We also measured the \ha\ excess after masking the central region that includes the $R$$_{50}$ radius to examine the \ha\ excess ratio in the outskirts of the galaxy. The mean $R$$_{50}$ of main and control galaxies are about 13.4 and 12.7 arcsec, respectively. Figure ~\ref{comp} shows distributions of the one-sided \ha\ excess ratios for brighter and fainter companions, separately measured for the entire galaxy (left) and the region beyond $R$$_{50}$ (right), respectively. Intriguingly, the distribution of \ha\ excess ratios for fainter companion galaxies (UGC00312NED01, CGCG536-030, NGC5480, and NGC3896) in the main sample exhibits a stark difference compared to the distribution for brighter companion galaxies. Moreover, fainter companion galaxies are significantly more likely to exhibit \ha\ excess ratios that deviate beyond 2$\sigma$ from the mean \ha\ excess ratio observed in isolated galaxies. Conversely, brighter companion galaxies, with the exception of one galaxy (UGC00312), tend to have \ha\ excess ratios that fall within 2$\sigma$ of the mean \ha\ excess ratio for isolated galaxies. This result reveals a significantly higher one-sided \ha\ excess ratios at the regions beyond $R$$_{50}$ of the galaxies compared to the values obtained for the entire galactic disks. Notably, three out of the four fainter companion galaxies exhibit a significantly higher likelihood of possessing one-sided \ha\ excess ratios deviating beyond 3$\sigma$ from the mean \ha\ excess ratio observed in isolated galaxies. As evident in Figure ~\ref{SDSS}, the sizes of the galaxies vary. Therefore, we calculated the \ha\ excess ratios for all sample galaxies by applying the $R$$_{90}$ region uniformly, and we note that the values do not differ significantly within two decimal places. These findings are suggestive of a potential influence of relative mass of companion on tidally induced star formation. Specifically, the outer regions of fainter companions may exhibit an enhanced sensitivity to the effects of tidal forces due to their relatively lower potential wells.  \

\begin{figure}
\centering
\includegraphics[width=8cm,height=7cm]{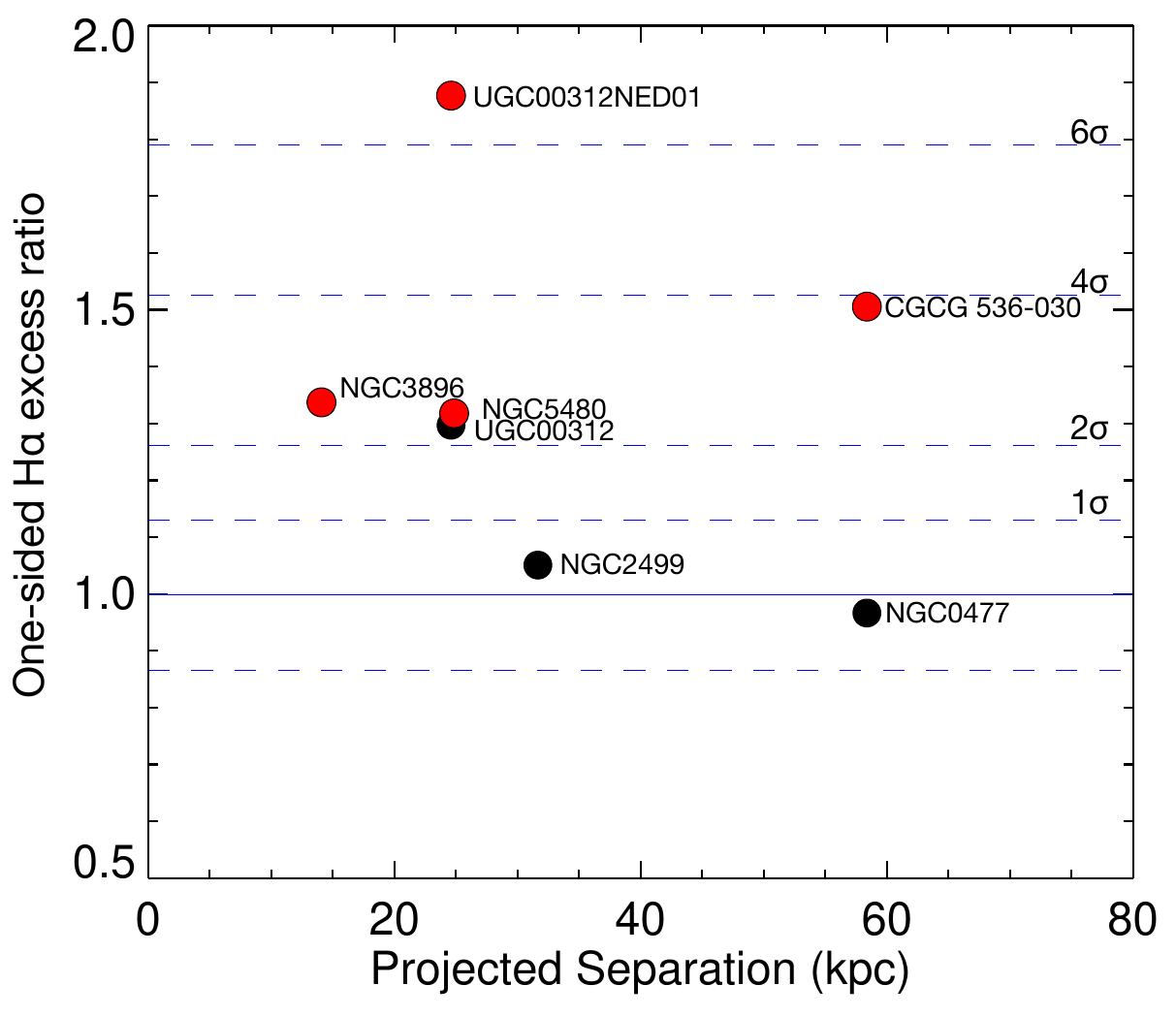}
\caption{The distribution of one-sided \ha\ excess ratios of main galaxies according to projected separation between pair galaxies. Symbols and lines are the same as those in Figure ~\ref{comp}. }
\label{sepa}
\end{figure}

Further supporting the above interpretation, previous research has established a general trend of enhanced star formation in galaxies as their projected separation decreases \citep{Barton00,Alonso07,Domingue09,Wong11,Ellison13}. Figure ~\ref{sepa} shows one-sided \ha\ excess ratios vesus projected separation. The observed relationship between \ha\ excess and separation exhibits a tendency towards a slight decrease in \ha\ excess with increasing separation. The limited sample size and the influence of star formation activity on a scale of up to 150 kpc \citep{Patton20} could be obscuring the detection of significant \ha\ excess differences within the narrow 60 kpc region. On the other hand, fainter galaxies tend to have higher \ha\ excess ratios even at the given separation. Intriguingly, within the main sample, cases where both galaxies in a pair were observed (UG00312-UGC00312NED01 and NGC0477-CGCG536-030) revealed that fainter galaxies exhibited a more pronounced increase in one-sided \ha\ excess ratio compared to their brighter counterparts. This also suggests a potential link between the stronger tidal forces experienced by fainter companions, as highlighted by our findings, and the observed one-sided \ha\ excess. The differential impact of tidal forces based on companion brightness could be a contributing factor to this general trend, where the increased susceptibility of fainter companions to tidal disturbances might lead to more pronounced star formation enhancements compared to their brighter counterparts, potentially explaining the observed one-sided \ha\ excess.

Some studies revealed an association between the spatial extent and level of interaction-triggered star formation, with each occurring on distinct timescales intrinsically linked to the evolutionary stage of merger \citep{CF2017MNRAS,CF2017AA,CF2017AAA}. Moreover, extended star formation was observed in some galaxies at the first pericentre passage \citep{Wang04,Elmegreen06}. We interpret these results as a natural extension of the extended star formation observed towards the companion galaxy in our main sample, serving as a subsequent stage of evolution for interacting systems. Another possible scenario for star formation enhancement in the galaxy outskirts could be ram pressure compression from the hot halo of a massive companion galaxy as the smaller companion approaches pericentre. Concurrently, gas stripping on the opposite side of the galaxy may occur. While our current sample does not exhibit any definitive signatures of gas stripping, further investigation through HI observations would be valuable to explore this possibility.

\

\section{Summary and Conclusion}

In this study, we investigated the characteristics of the \ha\ flux distribution in the phase before the first pericentre passage based on the eCALIFA pair sample, which covers the entire galaxy. The main results are as follows:

1) In order to examine the general characteristics of galaxies presumed to be in the phase before the first pericentre passage of galaxy interactions, we employed a selection criterion based on the absence of disturbance or tidal tails in optical images and the inclusion of the R$_{90}$ radius of galaxies within the eCALIFA FoV (on average, $\sim$1.3 R$_{90}$ of main galaxies). Additionally, isolated galaxies were chosen to serve as a control sample.

2) Intriguingly, \ha\ maps of galaxy pairs reveal elevated \ha\ flux in regions directly facing each other. Therefore, we measured the \ha\ excess ratio within the region facing the companion. Our findings demonstrate that fainter companion galaxies exhibit a significantly higher propensity to display \ha\ excess ratios exceeding 2$\sigma$ deviations from the mean \ha\ excess observed in isolated galaxies. Interestingly, our analysis revealed a heightened prominence of \ha\ excess at the outer regions ($>$$R$$_{50}$) of the galaxies. A significantly higher fraction (3 out of 4) of the fainter companion galaxies exhibit one-sided \ha\ excess ratios deviating beyond 3$\sigma$ from the mean value observed in isolated galaxies. This suggests that the mutually induced tidal force in interacting pairs manifests more prominently in outskirts of fainter galaxies due to their lower potential wells.

3) We further examined the variation in one-sided \ha\ excess ratio with projected separation, revealing a general trend of increasing one-sided \ha\ excess ratio as projected separation decreases. In line with the expected sensitivity to tidal effects, fainter galaxies exhibited higher one-sided \ha\ excess ratios at smaller separations.

Interactions between galaxies demonstrably play a crucial role in understanding galaxy evolution. Nonetheless, prior to this study, the properties of galaxies during the phase before the first pericentre passage were largely assumed to be akin to those of isolated galaxies, hence receiving minimal attention in simulations and observational studies in this phase. One promising avenue for advancing our understanding lies in the realm of spatially resolved high-resolution simulations. Previous simulation studies \citep{Teyssier10,Hopkins13,Renaud15,Renaud16} have yielded invaluable insights into the star formation efficiency outer region of galaxy during  galaxy mergers. On the other hand, statistical analyses conducted on expanded samples of early-phase interacting galaxy pairs are indispensable for solidifying the observed trends and potentially exposing hitherto unseen subtleties. A synergistic integration of high-resolution simulations and statistically robust analyses on larger samples of interacting pairs holds immense promise for unveiling a comprehensive picture of the fascinating early stages of galaxy interactions and their profound influence on galaxy evolution.

\section*{Acknowledgments}
We are grateful to the anonymous referee for helpful comments and suggestions that improved the clarity and quality of this paper. J.C. acknowledges support from the Basic Science Research Program through the National Research Foundation (NRF) of Korea (2018R1A6A3A01013232, 2022R1F1A1072874). J.H.L. $\&$ H.J. acknowledge support from the Basic Science Research Program through the National Research Foundation (NRF) of Korea (2022R1A2C1004025, $\&$ 2019R1F1A1041086, respectively). This work was supported by the Korea Astronomy and Space Science Institute under the R$\&$D program (Project No. 2024-1-831-01) supervised by the Ministry of Science and ICT (MSIT).

\section*{DATA AVAILABILITY}
This article is based on publicly available data from the eCALIFA \url{(http://ifs.astroscu.unam.mx/CALIFA_WEB/public_html/)}.




\bsp	
\label{lastpage}
\end{document}